# An approach for the automated risk assessment of structural differences between spreadsheets (DiffXL)

John Hunt
Excel for Managers
John.Hunt (at) ExcelForManagers.com

## ABSTRACT

This paper outlines an approach to manage and quantify the risks associated with changes made to spreadsheets. The methodology focuses on structural differences between spreadsheets and suggests a technique by which a risk analysis can be achieved in an automated environment. The paper offers an example that demonstrates how contiguous ranges of data can be mapped into a generic list of formulae, data and metadata. The example then shows that comparison of these generic lists can establish the structural differences between spreadsheets and quantify the level of risk that each change has introduced. Lastly the benefits, drawbacks and limitations of the technique are discussed in a commercial context.

## 1  INTRODUCTION

In business, a primary reason for choosing to use a spreadsheet over a fully managed computer application is the spreadsheet's flexibility: - Its ability to be updated or adapted quickly. Actual figures, for example monthly sales data, are often expected to change, but the calculation or formulae in a spreadsheet may also need regular review or up-dating to meet new requirements, legislation [1] or even the ability to correctly adapt to a new financial year.

In many organisations, changes to the workings and calculations of a spreadsheet are left relatively unmanaged. A requirement to create new documentation [2] or test the changes is often left to the conscience of the spreadsheet owner [3] [4]. Consequently managers, operational risk managers or auditors are frequently left unaware that a modification has even been made.

It follows that in many instances a change in spreadsheet cell value cannot be accurately attributed to an expected numerical change (say sales data) or a change caused by a difference in functionality … or both. [5]

The ability to make unregulated changes to a spreadsheet gives rise to a business and audit risk that is un-quantified and possibly, simply ignored.

In short, a spreadsheet that was once well designed, documented, tested and implemented, may well become prone to significant and often unseen errors over time.

## 2 BACKGROUND

### 2.1 Definition: Structural Change

In the context of this paper, a structural change is defined as the addition, removal, or modification of any formula or data in a worksheet cell. A structural change does NOT include a change to the format of a cell or a change in the displayed/calculated value of a previously defined formula cell.

### 2.2 Why test for structural changes?

Many spreadsheets used for reporting or valuation purposes are re-run on a cyclical basis. Good business practice dictates that any change to a spreadsheet's process or calculation method is documented and, where appropriate, suitable re-testing and/or approval carried out. However it is not unusual for undocumented and untested changes to creep into spreadsheets, often accidentally. A typical example of an accidental change is the replacement of a formula with an absolute value to mitigate a known error in source data. If the spreadsheet is subsequently saved, a structural change has been made, and future values have a high risk of being incorrect.

### 2.3 Minimising risk & semi-automated testing

Post-design, there are several simple ways in which businesses can help minimise the risks from spreadsheet changes. Some of these are physical methods, such as locking formula cells and protecting worksheet structure. Others are process driven such as a peer review of changes, the requirement to run pre-defined test scripts after each change, or using a spreadsheet comparison program to highlight value differences from identical input data.

### 2.4 Comparing spreadsheets: Formula view

A standard and useful test/auditing method is to show the formula view of a spreadsheet. The visibility of text or values entered directly (such as in B2:B4 below) is maintained, whilst the formulae themselves are shown (such as in D2) instead of their resultant value.

Reviewing formulae and values in Excel™ 2007

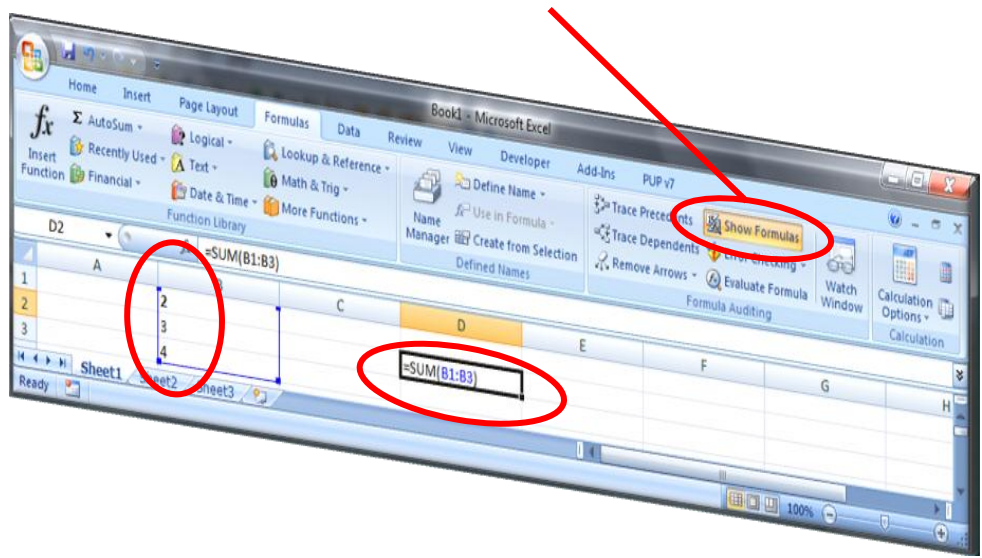

A key reason for viewing spreadsheets using this method is that direct structural spreadsheet comparisons can be made.

Given that there are no other differences between the spreadsheets (such as the use of different macros or named range changes etc.), it is safe to conclude that for any given set of inputs two spreadsheets with identical cell-for-cell formula views will give the same results: - Many spreadsheet comparison programs use a coded version of 'Formula View' as their primary basis of comparison.

**2.5  Auditing, testing & reviewing with a formula map**

Formula mapping programs are auditing tools that give a visual representation of the formula and data structure of a worksheet. They can be very useful when used after initial spreadsheet design or after any known changes have been made. They make it relatively simple to spot unexpected formulae variations or inconsistencies such as the accidental overwriting of a formula cell with a value.

Formula maps generally split a spreadsheet range into elements. In the example below, cells of the same colour represent cells with the same R1C1 formula, and static, directly entered values are shown in grey, making it relatively easy to spot some types of potential issue.

# 3 A SUGGESTED APPROACH FOR THE AUTOMATED RISK ASSESMENT OF SPREADSHEET CHANGES

## 3.1 Definition: DiffXL

DiffXL is the creation and subsequent comparison of contiguous range data and metadata, with a view to providing a risk based analysis of structural spreadsheet differences.

## 3.2 Creating Generic DiffXL Data Lists

This section proposes a method that can be used to ascertain if there are certain differences between spreadsheets or spreadsheet versions, and attempts to categorize the relative level of risk that the change has created.

*Please note that the terminology, samples and references within this paper relate to a Microsoft Excel™ implementation of the methodology. Although the terminology may differ, the principles should be identical in other spreadsheet applications.*

The proposed method first involves the storage of generic data from contiguous-range elements within each worksheet. Each element is deemed to consist of a rectangular range of cells with the same formulae (in R1C1 reference style) or a rectangular range of cells with *no* formula. The elements' details are stored for future comparative analysis. The actual data stored about these elements may take many forms, each with subtly differing analysis requirements and conclusions available. One such structure is described below by way of an example.

Consider the following mini-spreadsheet, shown below in normal, formula map and R1C1 formula view. The steps to create the rectangular element list are as follows …

|   | A | B | C | D | E |
|---|---|---|---|---|---|
| 1 | Base | X | Y | Gamma | Delta |
| 2 | CORONA | 7.031065 | 4.502179 | 0.01311 | 0.059024 |
| 3 | CORONA | 61.91627 | 17.35029 | 0.308808 | 5.357904 |
| 4 | CORONA | 126.6281 | 96.10542 | 4.662414 | 448.0832 |
| 5 | CORONA | 0.856899 | 0.316199 | 0.032906 | 0.010405 |
| 6 | CORONA | 14.93735 | 7.117257 | 0.428304 | 3.048348 |

|   | 1 | 2 | 3 | 4 | 5 |
|---|---|---|---|---|---|
| 1 | Base | X | Y | Gamma | Delta |
| 2 | CORONA | =Data!RC | =Data!RC | =Data!RC*Freq | =RC[-2]*RC[-1] |
| 3 | CORONA | =Data!RC | =Data!RC | =Data!RC*Freq | =RC[-2]*RC[-1] |
| 4 | CORONA | =Data!RC | 96.10542 | =Data!RC*Freq | =RC[-2]*RC[-1] |
| 5 | CORONA | =Data!RC | =Data!RC | =Data!RC*Freq | =RC[-2]*RC[-1] |
| 6 | CORONA | =Data!RC | =Data!RC | =Data!RC*Freq | =RC[-2]*RC[-1] |

The first 'rectangular' element is established. In this example it consists of six cells of non-formula data, highlighted. Note that although the first row is evidently a title, and rows 2 to 6 are other entries, they are both considered as a single rectangular element. The start row and column are recorded, along with rectangle's size, recorded as the number of rows and number of columns. There are no formulae for this rectangle, so the letter 'V' (for Value) is entered and the formula column is left blank.

(R1C1 reference view of the spreadsheet)         (Generic list of metadata)

| Row | Col | Rows | Cols | F/V | Formula |
|---|---|---|---|---|---|
| 1 | 1 | 6 | 1 | V | |
| 1 | 2 | 1 | 4 | V | |
| 2 | 2 | 5 | 1 | F | =Data!RC |
| 2 | 3 | 2 | 1 | F | =Data!RC |
| 2 | 4 | 5 | 1 | F | =Data!RC*Freq |
| 2 | 5 | 5 | 1 | F | =RC[-2]*RC[-1] |
| 4 | 3 | 1 | 1 | V | |
| 5 | 3 | 2 | 1 | F | =Data!RC |

The first rectangle element is now excluded, and the next rectangle is established. In this case it is the remainder of the title row. Its position and size is noted as before.

Note that in this particular example the rectangles are being established using an algorithm that broadly creates segments top to bottom and *then* left to right. The order in which the segments of any contiguous range are established is unimportant, but the method must be consistent for future comparisons to be accurate.

| Row | Col | Rows | Cols | F/V | Formula |
|---|---|---|---|---|---|
| 1 | 1 | 6 | 1 | V | |
| 1 | 2 | 1 | 4 | V | |
| 2 | 2 | 5 | 1 | F | =Data!RC |
| 2 | 3 | 2 | 1 | F | =Data!RC |
| 2 | 4 | 5 | 1 | F | =Data!RC*Freq |
| 2 | 5 | 5 | 1 | F | =RC[-2]*RC[-1] |
| 4 | 3 | 1 | 1 | V | |
| 5 | 3 | 2 | 1 | F | =Data!RC |

The third rectangle element is a series of 5 formulae, whose R1C1 formulae are identical. The position and size is recorded. An 'F' (for formula) is noted, along with the R1C1 formula.

| Row | Col | Rows | Cols | F/V | Formula |
|---|---|---|---|---|---|
| 1 | 1 | 6 | 1 | V | |
| 1 | 2 | 1 | 4 | V | |
| 2 | 2 | 5 | 1 | F | =Data!RC |
| 2 | 3 | 2 | 1 | F | =Data!RC |
| 2 | 4 | 5 | 1 | F | =Data!RC*Freq |
| 2 | 5 | 5 | 1 | F | =RC[-2]*RC[-1] |
| 4 | 3 | 1 | 1 | V | |
| 5 | 3 | 2 | 1 | F | =Data!RC |

The 5 remaining elements are calculated and recorded. Note the inconsistent value in row 4, column 3 which prevented a single large area (R2C2 to R6C3) being created.

| Row | Col | Rows | Cols | F/V | Formula |
|---|---|---|---|---|---|
| 1 | 1 | 6 | 1 | V | |
| 1 | 2 | 1 | 4 | V | |
| 2 | 2 | 5 | 1 | F | =Data!RC |
| 2 | 3 | 2 | 1 | F | =Data!RC |
| 2 | 4 | 5 | 1 | F | =Data!RC*Freq |
| 2 | 5 | 5 | 1 | F | =RC[-2]*RC[-1] |
| 4 | 3 | 1 | 1 | V | |
| 5 | 3 | 2 | 1 | F | =Data!RC |

### 3.3 Recording static information

In order to ascertain if any change between spreadsheets has been made, there is also a requirement to assess any change in static information. To give a guaranteed 100% accuracy then both spreadsheets must be opened and a cell by cell comparison made, or every static cell's location and value must be stored. Assuming the preferred option is to store the values in the generic list, this can be achieved as follows…

By definition, each rectangular segment (line in the list) must consist of either formulae or static data. For each segment that contains static data, concatenate the cells' string values and separate *every* cell with a delimiting character (for example '|') irrespective of the cell's status. In the previous example, the first segment could therefore be made to include the data string…

*Base|CORONA|CORONA|CORONA|CORONA|CORONA* and this string could be stored in the Formula column, I.e.

| | 1 | 2 | 3 | 4 | 5 |
|---|---|---|---|---|---|
| 1 | Base | X | Y | Gamma | Delta |
| 2 | CORONA | =Data!RC | =Data!RC | =Data!RC*Freq | =RC[-2]*RC[-1] |
| 3 | CORONA | =Data!RC | =Data!RC | =Data!RC*Freq | =RC[-2]*RC[-1] |
| 4 | CORONA | =Data!RC | 96.105 | =Data!RC*Freq | =RC[-2]*RC[-1] |
| 5 | CORONA | =Data!RC | =Data!RC | =Data!RC*Freq | =RC[-2]*RC[-1] |
| 6 | CORONA | =Data!RC | =Data!RC | =Data!RC*Freq | =RC[-2]*RC[-1] |

| Row | Col | Rows | Cols | F/V | Formula |
|---|---|---|---|---|---|
| 1 | 1 | 6 | 1 | V | Base|CORONA|CORONA|COR |
| 1 | 2 | 1 | 4 | V | |
| 2 | 2 | 5 | 1 | F | =Data!RC |
| 2 | 3 | 2 | 1 | F | =Data!RC |
| 2 | 4 | 5 | 1 | F | =Data!RC*Freq |
| 2 | 5 | 5 | 1 | F | =RC[-2]*RC[-1] |
| 4 | 3 | 1 | 1 | V | |

This exacting method may be appropriate or essential in some circumstances. However it can become cumbersome for large segments, and result in generic list file sizes approaching or even exceeding that of the original spreadsheet.

One alternative is to concatenate the cell values as above, but hold a checksum based on the entire string. Although it would not be possible to ascertain which cell(s) had changed directly from the checksum, it would still be possible to ascertain with a very high degree of probability that a change had been made within any data segment, and if necessary the two files opened to retrieve the difference by cell. The commercial implementation of the methodology uses this method and a CRC-32 checksum with 2^32 possible values as a default setting.

A small but important limitation for storing static values in this way is that any cells containing a value consisting solely of the separating character or characters (e.g. ||| in the above example) could fool a change-test algorithm. In the commercial implementation of this method, a non-printing character is used as the separating character to help minimise this risk.

### 3.4 Additional requirements

Two additional fields are also required in order to store the best level of information for future analysis:- Firstly the worksheet *object* name – this name enables us to compare worksheets, even if the user has changed the name on the worksheet tab. Secondly, an arbitrary index of the contiguous range: - Although strictly speaking, contiguous range locations and sizes can be calculated from the information contained in the other fields, this additional value assists in their quick identification and simplifies some of the required analysis.

A final example of a segment list for a worksheet with two contiguous ranges is show below. Note the addition of the two extra columns, and in this example the 8 character representation of the CRC-32 checksums used for value entries.

| Sheet | Contig ID | Row | Col | Rows | Cols | F/V | Formula |
|---|---|---|---|---|---|---|---|
| Sheet1 | 1 | 1 | 1 | 6 | 1 | V | CAA8E7F2 |
| Sheet1 | 1 | 1 | 2 | 1 | 4 | V | 1CE2BE24 |
| Sheet1 | 1 | 2 | 2 | 5 | 1 | F | =Data!RC |
| Sheet1 | 1 | 2 | 3 | 2 | 1 | F | =Data!RC |
| Sheet1 | 1 | 2 | 4 | 5 | 1 | F | =Data!RC*Freq |
| Sheet1 | 1 | 2 | 5 | 5 | 1 | F | =RC[-2]*RC[-1] |
| Sheet1 | 1 | 4 | 3 | 1 | 1 | V | DF578763 |
| Sheet1 | 1 | 5 | 3 | 2 | 1 | F | =Data!RC |
| Sheet1 | 2 | 8 | 2 | 1 | 4 | F | =AVERAGE(R[-6]C:R[-2]C) |

### 3.5 Drawing risk based conclusions from the comparison of two DiffXL lists

In order to draw appropriate conclusions, a comparison of two lists is required at worksheet, contiguous range or element level. Both qualitative and quantitative conclusions can be made.

At this stage, it is also possible to include a subjective risk assessment of any changes. The actual and genuine level of risk is dependent on numerous factors, including the type of data, type of change, and number of simultaneous changes.

The *category* of risk also needs to be taken into account. For example risk may be considered in different terms if the user is looking at the risk of an incorrect result, the risk of a catastrophic malfunction of the spreadsheet or the risk of failing an audit, and so on. A variety of change-test algorithms, or settings within the algorithms would therefore be deemed to be appropriate within different industries or departments.

Below are some of the types of change to spreadsheet structure that can be ascertained using DiffXL, together with their suggested (user-configurable) associated level of risk.

| Comparison | Risk |
|---|---|
| Entire range moved | Low |
| New static text, placed away from other ranges | Low |
| One or more ranges sorted | Low |
| Text or headers changed | Medium |
| Contiguous range split into two or more segments | Medium |
| New items added within a range | Medium |
| Formulae change throughout an entire range | Medium |
| Items removed from within a range | Medium |
| Items removed or changed, breaking up a range | Medium |
| Numeric cell changed to a number formatted as text | High |
| Overwrite a formula with a number or text | High |
| New data and formulae introduced | High |
| New or deleted worksheets | High |

# 4 BENEFITS, DRAWBACKS & LIMITATIONS OF USING THE DiffXL METHODOLOGY

## 4.1 Benefits of using the DiffXL methodology

The ability to demonstrate that no structural differences have been made to a spreadsheet over time, or to be able to identify where changes have been made may be of benefit to spreadsheet owners, managers and auditors in many commercial organisations.

DiffXL's ability to assign a relatively low risk indicator to simple changes such as the movement of a block of cells from the top to the bottom of a spreadsheet, or the sorting of a data range, means testing and approval effort can be best allocated where changes have been identified as more likely to result in material differences to calculated values.

Once a DiffXL list has been created for a spreadsheet, that list is generally the only requirement to perform difference tests or analysis. If the checksum method for storing static values is used, there are few security issues with releasing the lists for external analysis or reporting, even from within strictly secure or sensitive environments.

If run in the background, either directly when a spreadsheet change is made or at a pre-defined frequency, owners or managers can be alerted only if a change of a certain magnitude of risk is made. This offers the potential for a very full, discreet and un-obtrusive testing regime from the user's viewpoint. [6]

## 4.2 Drawbacks & limitations of using the DiffXL methodology

Changing spreadsheet data or formulae is only one way to change the operation of a spreadsheet. In a commercial application, other aspects of the spreadsheet also need to be considered and analysed, for example changes to macros, named ranges or data connections.

Unless direct cell-to-cell comparison between spreadsheets for static values is used, there is a small risk that a value change will not be detected by this methodology. In most implementations the risk is negligible, but nevertheless possible and dependent on the method in which the static data is stored (directly or through a checksum), the quality of any checksum, and the use of the separating character.

Although based on objective algorithms, overall risk assessments may still be subjective. As an example, should three low risk changes constitute a low, medium or high risk change?

## 5 DiffXL IN COMMERCIAL APPLICATIONS

In a commercial risk management product, Diff XL is likely to be combined with more established spreadsheet metric and direct cell comparison testing methods.

Listed below are screenshots from two products where the methodology has been combined in this way, and which highlight areas in which DiffXL could be applied.

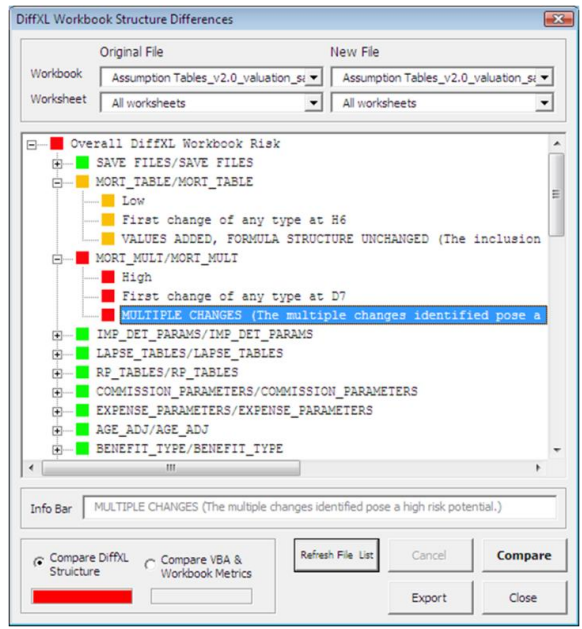

a) An auditor's tool that directly compares spreadsheets from one year to the next, to ascertain the level of risk associated with all the changes. No software installation is required by the auditor as the tool can be run directly from a memory stick if required.

b) A SharePoint™ Control Framework that monitors spreadsheets and alerts the owner if cumulative changes have exceeded a pre-defined level of risk. Here, the spreadsheet's complexity and change-risk at each 'publish' is charted for the owner or auditor.

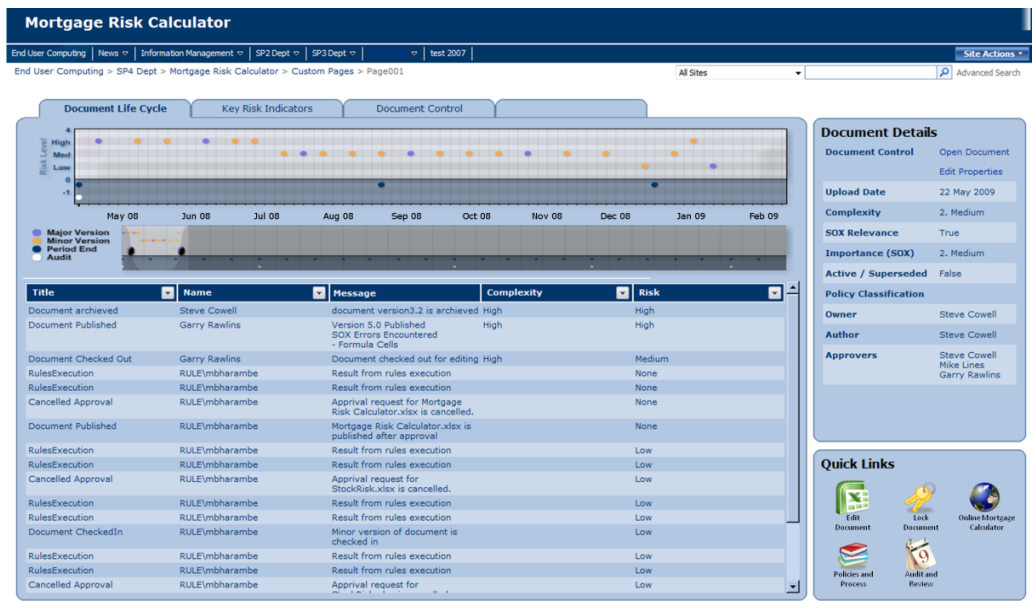

Further details of DiffXL Audit and the IDAS SharePoint Control Framework can be obtained from the author.

## 6 ACKNOWLEDGEMENTS

The author wishes to express thanks to Mike Lines of Rule Financial Ltd for his assistance with the commercial implementation of DiffXL.

## 7 SUMMARY

DiffXL is an alternative methodology to compare spreadsheets.

DiffXL does not compare spreadsheets directly and is not a substitute or replacement for any of the established auditing and comparison methods or tools.

For some owners, risk managers or auditors it may be relatively more important for them to understand the risk profile of the difference rather than the difference per se, and for these people DiffXL may offer insight and useful information.

DiffXL may be of particular interest to auditors and companies that desire discreet ongoing spreadsheet control management.

## 8 REFERENCES

(1) PricewaterhouseCoopers., The Use of Spreadsheets: Considerations for Section 404 of the Sarbanes-Oxley Act. 2004

(2) Automating Spreadsheet Discovery & Risk Assessment. Eric Perry, 2008, http://arxiv.org/abs/0809.3016

(3) Reducing Overconfidence in Spreadsheet Development. Dr. Raymond R. Panko, University of Hawaii (2008). http://arxiv.org/abs/0804.0941

(4) Facing the Facts. Patrick O'Beirne, Systems Modelling. 2008. http://arxiv.org/abs/0803.3394

(5) Impact of Errors in Operational Spreadsheets. Stephen G. Powell, Barry Lawson, Kenneth R. Baker 2007. http://arxiv.org/abs/0801.0715

(6) Risk Assessment For Spreadsheet Developments: Choosing Which Models to Audit. Raymond J. Butler, 2008. http://arxiv.org/abs/0805.4236